\documentstyle[amssymb,aps,multicol,epsf]{revtex}

\begin{document}
\draft

\title{
How to construct a correlated net
}

\author{
S.N. Dorogovtsev$^{1, 2, \ast
}$,  
J.F.F. Mendes$^{1, \dagger}$, 
and 
A.N. Samukhin$^{2, \ddagger}$
}

\address{
$^{1}$ Departamento de F\'\i sica and Centro de F\'\i sica do Porto, Faculdade 
de Ci\^encias, 
Universidade do Porto\\
Rua do Campo Alegre 687, 4169-007 Porto, Portugal\\
$^{2}$ A.F. Ioffe Physico-Technical Institute, 194021 St. Petersburg, Russia 
}

\maketitle
   
\begin{abstract} 
(a) We propose a ``static'' construction procedure for random networks with given correlations of the degrees of the nearest-neighbor vertices. This is an equilibrium graph, maximally random under the constraint that its degree-degree distribution is fixed. 
(b) We generalize the notion of preferential linking and introduce a new category, {\em pair preference} and a pair preference function for the attaching of edges to pairs of vertices. This allows dynamically generate equilibrium correlated networks.      
\end{abstract}


\begin{multicols}{2}
\narrowtext


The basis of the network science (see recent reviews \cite{s01,ab01a,dm01c}) 
are construction procedures. There are four main construction procedures of networks (and a large number of their versions and variations): 

\begin{itemize} 

\item[(i)] 
The classical random graphs of Erd\"{o}s and R\'enyi \cite{er59}, which are graphs with a Poisson degree distribution obtained by the random connection of their vertices. 

\item[(ii)] The random graphs with a given degree distribution, or, more rigorously, the labeled random graphs with a given degree sequence, which are also called {\em the configuration model}, see Refs. \cite{mr95}. These are the graphs, maximally random under the constraint that their degree distribution is a given function. Dynamical constructions for these graphs, incorporating the mechanism of preferential linking, are discussed in Refs. \cite{bck01,dms02c}.   

\item[(iii)] Small-world networks, that is, the superposition of the regular lattices and the classical random graphs, which were introduced by Watts and Strogatz \cite{ws98} and combine the compactness of the classical random graphs and the strong clustering of many regular lattices. 

\item[(iv)] Networks, growing under the mechanism of preferential linking (including, as a particular case, random linking), which were introduced by Barab\'asi and Albert \cite{ba99}.  

\end{itemize}  

The simple types of correlations in networks are: 

\begin{itemize} 

\item[(1)] 
Pair degree-degree correlations, which are the correlations of the degrees of the nearest-neighbor vertices (see, e.g., Ref. \cite{kr00}). These correlations are described by the distribution of the degrees of the end vertices of a randomly chosen edge, $P(k,k')$. The absence of pair correlations means the following factorization: 

\begin{equation}
P(k,k^\prime) = \frac{kP(k) k^\prime P(k^\prime)}{ [\sum_k kP(k)]^2}
\, .  
\label{4.62}
\end{equation} 

\item[(2)] 
Loops, and, in particular, clustering, since the clustering coefficient of a network is actually ``the concentration'' of loops of the length three in a network. 

\end{itemize} 

Constructions (i) and (ii) generate uncorrelated graphs: degree-degree correlations and the relative number of loops approaches zero in the thermodynamic limit (the limit of an infinite network). 

Constructions (iii) and (iv), as a rule, produce correlated networks. Construction (iii) leads to a finite clustering even in the thermodynamic limit. Construction (iv) effectively generates pair correlations. 

Here we briefly describe natural constructions of equilibrium correlated graphs. 

{\em Static (geometric) construction}.---This is the direct generalization of construction (ii): a construction procedure for a graph with a given degree-degree distribution for nearest neighbors, $P(k,k^\prime)$, and a fixed number of vertices, $N$. In the spirit of the configuration model, this is the statistical ensemble of random graphs, which contains all graphs with a given $P(k,k^\prime)$, where each realization is taken with equal statistical weight. In Ref. \cite{dms02c} we called such constructions a microcanonical ensemble of random graphs.  

How can such graphs be constructed for large $N$? 
Notice that in these networks, the ordinary degree distribution follows from the degree-degree distribution: 

\begin{equation} 
\sum_k P(k,k^\prime) = \frac{kP(k)}{\sum_k kP(k)} 
\, .  
\label{4.63}
\end{equation} 
This relation gives the function $P(k)$ up to a constant factor, 
which, in turn, can be obtained from the normalization condition $\sum_k P(k) = 1$. Thus we get 
\begin{equation} 
\overline{k} = \left[\sum_{k,k^\prime}P(k,k^\prime)/k\right]^{-1}
\, ,  
\label{4.63}
\end{equation} 
and so the total number of edges is known, $L=\overline{k}N/2$. 

So, we have $N$ and $L$ fixed and can formulate the construction procedure as it follows: 

\begin{itemize}

\item[(a)] Using the given $P(k,k^\prime)$ find the total number of edges, $L$. 

\item[(b)] Create the infinite number $L$ of pairs of integers $(k,k^\prime)$ distributed as $P(k,k^\prime)$. These are $L$ edges with ends of degrees $k$ and $k^\prime$. Therefore we have edges labeled by degrees of their ends but yet have no vertices. 

\item[(c)] Select, at random, groups of ends of degree $k$, 
each one consisting of $k$ ends. 
Tie them in bunches, each of $k$ tails, that is, attach them to vertices of degree $k$. 

\end{itemize}
One can check that this procedures is possible in the thermodynamic limit, and it generates tree-like (locally) graphs with the relatively low number of loops like construction (ii).  

{\em Dynamical construction}.---We can also 
construct such graphs dynamically. 
For example, this can be made by applying the following process on a graph. 
Randomly selected edges are removed to new positions chosen preferentially, that is, to pairs of vertices chosen with a preference $f(k,k^\prime)$. In the long-time limit, we obtain an equilibrium network with degree-degree correlations.

Here we introduce a new category, {\em pair preference} and a pair preference function. This is a more general notion than a one-vertex preference, where the probability that the edge becomes attached, $f(k)$, depends only on the degree of a vertex. Indeed, an edge has not one but two ends. In general, they must become attached not to one but to two vertices, and so it is a combination of two degrees that is natural form of preference \cite{remark4}.

In this dynamical construction (in fact, this is a canonical ensemble of correlated random networks, see Ref. \cite{dms02c}), the given preference $f(k,k^\prime)$ and the total number of edges, $L$, determine $P(k,k^\prime)$\vspace{4pt}.

{\em Note added}.---Another construction of correlated networks, which is close to the above dynamical construction, was recently proposed  by Berg and L\"assig, Ref. \cite{bl02c}. 
\\

\noindent
$^{\ast}$      E-mail address: sdorogov@fc.up.pt \\
$^{\dagger}$   E-mail address: jfmendes@fc.up.pt \\ 
$^{\ddagger}$  E-mail address: alnis@samaln.ioffe.rssi.ru

\end{multicols}


\begin{references} 

\bibitem{s01}  S.H. Strogatz, Nature {\bf 401}, 268 (2001). 

\bibitem{ab01a}  R. Albert and A.-L. Barab\'{a}si, 
Rev. Mod. Phys. {\bf 74}, 47 (2002). 

\bibitem{dm01c}  S.N.~Dorogovtsev~and~J.F.F.~Mendes, 
Adv. Phys. {\bf 51}, 1079 (2002). 

\bibitem{er59}  P. Erd\"{o}s and A. R\'enyi,  
Publications Mathematicae {\bf 6}, 290 (1959); 
Publ. Math. Inst. Hung. Acad. Sci. {\bf 5}, 17 (1960). 

\bibitem{mr95}  
A. Bekessy, P. Bekessy, and J. Komlos, Stud. Sci. Math. Hungar. {\bf 7}, 
343 (1972); 
E.A. Bender and E.R. Canfield, J. Combinatorial Theory A {\bf 24}, 296 (1978); 
B. Ballob\'{a}s, Eur. J. Comb. {\bf 1}, 311 (1980); 
N.C. Wormald, J. Combinatorial Theory B {\bf 31}, 156,168 (1981);
M. Molloy and B. Reed, Random Structures and Algorithms {\bf %
6}, 161 (1995).

\bibitem{bck01}  Z. Burda, J. D. Correia, and A. Krzywicki, 
Phys. Rev. E {\bf 64}, 046118 (2001). 

\bibitem{dms02c} 
S.N. Dorogovtsev, J.F.F. Mendes, and A.N. Samukhin, 
cond-mat/0204111. 

\bibitem{ws98} 
D.J. Watts and S.H. Strogatz,  
Nature {\bf 393}, 440 (1998).  

\bibitem{ba99}  A.-L. Barab\'{a}si and R. Albert, Science {\bf 286}, 509
(1999). 







\bibitem{kr00}  
P.L. Krapivsky and S. Redner,    
Phys. Rev. E {\bf 63}, 066123 (2001).  









\bibitem{remark4}
Preferential attachment was introduced by using a rather specific network, a citation graph: the Barab\'asi-Albert model \protect\cite{ba99}. In such graphs, new vertices become attached to old ones, and actually only one of the ends of the each edge becomes attached preferentially. In this situation the usual form of preference, $f(k)$, is the only possibility. 

\bibitem{bl02c} 
J. Berg and M. L\"assig, 
cond-mat/0205589. 


\end{references}
\end{document}